\documentclass [12pt, one column, a4paper] {article}

\usepackage{amsmath,amsfonts,amssymb}
\usepackage{bm,graphicx,graphics,color}

\title {External Sources in Lee-Wick Theories}

\author{F.A. Barone\thanks{e-mail: fbarone@unifei.edu.br}, G. Flores-Hidalgo\thanks{e-mail: gfloreshidalgo@unifei.edu.br}\\
{\small ICE - Universidade Federal de Itajub\'a, Av. BPS 1303}\\
{\small Caixa Postal 50 - 37500-903, Itajub\'a, MG, Brazil.}\\
A.A Nogueira\thanks{e-mail:nogueira@ift.unesp.br}\\
{\small IFT - R. Dr. Bento Teobaldo Ferraz, 271}\\
{\small Bairro: Barra-Funda, 01140-070, S\~ao Paulo, SP, Brazil.}}

\date {}

\begin {document}

\baselineskip=20pt

\maketitle

\begin{abstract}
We investigate some peculiar aspects of the so called Lee-Wick Electrodynamics focusing on physical effects produced by the presence of sources for the vector field. The interactions between stationary charges distributions along parallel branes with arbitrary dimensions is investigated and the energy of a point charge is discussed. Some physical phenomena produced in the vicinity of a Dirac string are also investigated. We consider the Lee-Wick theory for the scalar field, where it can emerge some interesting effects with no counterpart for the vector gauge field theory.
\end{abstract}

\section{Introduction}

From time to time models with derivatives of superior order in the field variables have been studied in the literature. As far as the authors know, one of the first models of this kind was proposed by B. Podolsky \cite{Podolsky42,Podolsky44,Podolsky48} and T. Lee and G. Wick \cite{LW69,LW70} and some of its technical features are the fact that, in 3+1 dimensions, the self energy for a point-like electric charge is finite, and it exhibits gauge invariance as well as a non vanishing pole for the gauge field propagator in momenta space. Nowadays we can find a vast literature about those kinds of models, where they are usually referred to as Lee-Wick models.

Recently, after the propose of the so called Lee-Wick Standard Model (LWSM) \cite{Grinstein2008}, some interest in Lee-Wick type models has been aroused in many contexts. We can cite, for instance, possible experimental signatures of LWSM as well as experimental and/or theoretical constraints on its parameters \cite{Espinosa2008,Underwood2009,CaronePRD2009,RizzoJHEP2008,RizzoJHEP2007,Schat2008,KraussPRD2008,CuzinattoIJMPA2011,AcciolyMPLA2010,AcciolyMPLA2011,Accioly2010}. Some effort has also been spend in theoretical aspects of the LWSM, as the treatment of the ghosts states \cite{Shalaby2009}, possible minimal LWSM \cite{CaronePLB2008}, models with higher derivatives \cite{CaronePLB2009,CaroneJHEP2009}, the renormalization of Lee-Wick gauge theories \cite{GrinsteinPRD2008,gc} and finite temperature Lee-Wick theories \cite{FornalPLB2009,BoninPRD2010,BoninPRD2011}.   

In what concerns  Lee-Wick type field theories with higher degrees of freedom, the study of non-Abelian Lee-Wick gauge theories \cite{Alekseev,BallNPB83,Baskal93,GOWPRD2008,GrinsteinPRD2008} and peculiarities of Lee-Wick type theories for gravity \cite{StellePRD77,StelleGRG77,WuPLB2008,WuPRD2008,Rodigast2009,Accioly2003} are subjects studied since a long time. In this context we would like to call attention to some cosmological implications of Lee-Wick scalar field \cite{Cai2009,ChoEPJC2013,ChoJCAP2011} and its use for describing the dark energy \cite{LeeIJMP2008}. 

Many other interesting studies of Lee-Wick theories can be found in the literature, among them we can cite, for instance, the fact that Lee-Wick electrodynamics also leads to a finite self force for a point-like electric charge \cite{Frankel,Zayats}, the study of waves propagation in Lee-Wick theories \cite{SantosMPLA2011}, the role of Lee-Wick models in dynamical breaking of chiral symmetry \cite{GabrielliPRD2008}, the quantization of Lee-Wick electrodynamics \cite{GalvaoCJP,Bufalocan,BufaloPRD83}, the connection between electrodynamics with minimal length and Lee-Wick electrodynamics \cite{Moyaedi}, theta term generalization in connection with non-commutative electrodynamics \cite{GaeteJPA2012}, formalism of second order gauge theories for Lee-Wick theories \cite{CuzinattoAP2007}, first order formalism for Lee-Wick electrodynamics \cite{KruglovJPA2010}, generalizations with auxiliary fields with higher derivatives \cite{ChoPRD2010} and so on.

One of the most fundamental questions one can make, about gauge field models with higher derivatives concerns on the physical phenomena produced by the presence of field sources, mainly on the phenomena with no counterpart in the standard theories with no superior derivatives \cite{AcciolyPRD2004}. This paper is devoted to this subject in the context of Lee-Wick Electrodynamics, where we search for effects produced by the presence of  field sources  not present in the Maxwell theory. For completeness we consider also the extension of the Lee-Wick Electrodynamics for the massive scalar field. This last model leads to interesting peculiarities once we can have two mass parameters.    

In section (\ref{secaocargas}) we  study the interaction between stationary sources for the vector field in the Lee-Wick Electrodynamics. The sources are taken to be distributed along parallel branes with arbitrary dimensions. We focus on the case where the sources describe uniform and stationary distributions of electric charges along the branes. For completeness, we also discuss the case of sources which describe uniform distributions of electric dipoles. From the computed results we investigate the behavior of the energy between two electric charges when the distance between them is small. The results indicate that, for spatial dimensions higher than 3, the self energy of a point-like electric charge diverges in the Lee-Wick Electrodynamics. In section (\ref{secaoDirac}) we investigate some peculiarities of Lee-Wick Electrodynamics (in comparison with the Maxwell theory) in the vicinity of one or two Dirac strings. In section (\ref{secaoescalar}) we consider the massive Lee-Wick scalar field, mainly in what concerns the interaction between point-like field sources. Once we can have two mass parameters, we can obtain results with no counterpart in comparison with Klein-Gordon theory. Section (\ref{secaoconc}) is dedicated for the conclusions and final remarks. 

Along this paper we work in Minkowsky space-time, with diagonal metric $\eta^{\mu\nu}=(1,-1,...,-1)$, $D+d$ spatial dimensions and one time dimension. The time coordinate shall be represented by $x^{0}$ and the $(D+d+1)$-vector position shall be designated by
\begin{equation}
\label{def4vetor}
x=(x^{0},x^{1},...,x^{d},x^{d+1},...,x^{d+D})\ .
\end{equation}
We shall also use the following notations for spatial coordinates perpendicular and parallel to the branes:
\begin{eqnarray}
\label{defxperpx|}
{\bf x}_{\perp}&=&(x^{1},...,x^{d})\nonumber\\
{\bf x}_{\|}&=&(x^{d+1},...,x^{d+D})\ ,
\end{eqnarray}
and similar ones for the momentums $p$.

\section{Charged stationary branes}
\label{secaocargas}

In this section we investigate some aspects of the interactions between stationary sources for the vector field in the Lee-Wick Electrodynamics in an arbitrary number of spatial dimensions. For simplicity, the sources are taken to be concentrated along stationary parallel branes. 

The abelian Lee-Wick Electrodynamics is described by the lagrangian density \cite{LW69,LW70}	
\begin{equation}
\label{defL}
{\cal L}=-\frac{1}{4}F_{\mu\nu}F^{\mu\nu}-\frac{1}{4m^{2}}F_{\mu\nu}\partial_{\alpha}{\partial}^{\alpha}F^{\mu\nu}-\frac{{(\partial_{\mu}A^{\mu})}^2}{2\xi}-J_{\mu}A^{\mu}\ ,
\end{equation}
where $J^{\mu}$ is the vector external source,
\begin{eqnarray}
F_{\mu\nu}&=&\partial_{\mu}A_{\nu}-\partial_{\nu}A_{\mu}
\end{eqnarray}
is the field strength, $A^{\mu}$ is the vector potential and $m$ is a parameter with mass dimension. The third term on the right hand side of (\ref{defL}) was introduced in order to fix the gauge and $\xi$ is a gauge fixing parameter.

Neglecting surface terms, one can write
\begin{equation}
{\cal L}=\frac{1}{2}A_{\mu}{\cal O}^{\mu\nu}A_{\nu}-J_{\mu}A^{\mu}
\end{equation}
where we defined the differential operator
\begin{equation}
\label{defO}
{\cal O}^{\mu\nu}={\eta}^{\mu\nu}\Biggl(1+\frac{\partial^{\beta}\partial_{\beta}}{m^{2}}\Biggr)\partial^{\alpha}\partial_{\alpha}
-\Biggl(1-\frac{1}{\xi}+\frac{\partial^{\beta}\partial_{\beta}}{m^{2}}\Biggr){\partial}^{\mu}{\partial}^{\nu}\ .
\end{equation}

The propagator $D^{\mu\nu}(x,y)$ is the inverse of the operator ${\cal O}^{\mu\nu}$ in the sense that
\begin{equation}
\label{defD}
{\cal O}^{\mu\nu}D_{\nu\lambda}(x,y)=\eta^{\mu}_{\ \lambda}\delta^{4}(x-y)\ .
\end{equation}

One can get the propagator by using standard field theory methods. Searching for a Fourier representation for $D_{\nu\lambda}(x,y)$ one can show that
\begin{equation}
\label{propagador}
D_{\mu\nu}(x,y)=\int \frac{{d^{d+D+1}p}}{(2\pi)^{d+D+1}}
\Biggl(\frac{1}{p^{2}-m^{2}}-\frac{1}{p^{2}}\Biggr)
\left\{{\eta}_{\mu\nu}-\frac{p_{\mu}p_{\nu}}{p^{2}}\biggl[1+\xi\biggl(\frac{p^{2}}{m^{2}}-1\biggr) \biggr]\right\}\exp \biggl[-ip(x-y)\biggr].
\end{equation}

Once we have a quadratic lagrangian in the field variables $A^{\mu}$, the energy of the system due to the presence of the sources is given by  \cite{Zee}
\begin{equation}
\label{Egeral}
E=\lim_{T\to\infty}\frac{1}{2T}\int d^{d+D+1}x\ d^{d+D+1}y\ J^{\mu}(x)D_{\mu\nu}(x,y)J^{\nu}(y)\ .
\end{equation}

As discussed in references \cite{BaroneHidalgo1,BaroneHidalgo2}, the presence of stationary uniform distributions of charges along $D$-dimensional parallel branes can be described by the external source
\begin{equation}
\label{defJ}
J_{\mu}=\eta_{\mu0}\sum^{N}_{k=1}\lambda_{k}\delta^{d}(\bf{x}_{\bot}-\bf{a}_{k})\ .
\end{equation}
In expression (\ref{defJ}), $N$ is the number of branes, $\lambda_{k}$ is the charge density for the $k$-th brane (charge per unity of brane area), ${\bf a}_{k}$ designates the $k$-th brane position and ${\bf x}_{\bot}$ are the coordinates perpendicular to the branes. It is important to notice that the vectors ${\bf a}_{k}$ have only perpendicular coordinates, {\it i.e.}, ${\bf a}_{k}=(a_{k}^{1},\dots,a_{k}^{d},0,0,\dots)$.   

Substituting the source (\ref{defJ}) into (\ref{Egeral}) and using the propagator (\ref{propagador}),  identifying  the area of a given brane $L^{D}=\int d^{D}{\bf y}_{\|}$, and neglecting the contributions  due to the self interactions of each brane with itself (the self energy of the branes), we have for the energy per unit brane area
\begin{equation}
\label{Eintermediario2}
{\cal E}=\frac{E}{L^{D}}=\sum^{N}_{k=1}{\sum^{N}_{s=1}}\lambda_{k}\lambda_{s}(1-\delta_{k,s})\frac{1}{2}
\Biggl(\int \frac{d^{d}{\bf p}_{\bot}}{(2\pi)^{d}}\frac{1}{{\bf p}_{\bot}^{2}}\exp(i{\bf p}_{\bot}.{\bf a}_{ks})
-
\int \frac{d^{d}{\bf p}_{\bot}}{(2\pi)^{d}}\frac{1}{{\bf p}_{\bot}^{2}+m^2}\exp(i{\bf p}_{\bot}.{\bf a}_{ks})\Biggr)
\ ,
\end{equation}
where ${\bf a}_{ks}={\bf a}_{k}-{\bf a}_{s}$

Notice that the energy splits into two contributions. The first one comes from the massless sector of the model, and does not involve the parameter $m$. The other one comes from the massive sector. The first contribution leads to a Coulomb-like interaction, as discussed in reference \cite{BaroneHidalgo1}, with repulsive behavior (charges with the same signal repel each other). The second contribution produces a Yukawa-like interaction with attractive behavior. One can see this fact by solving the integrals in (\ref{Eintermediario2}). For this task we can use the results of reference \cite{BaroneHidalgo1},
\begin{equation}
\label{intm}
\int \frac{d^{d}{\bf p}_{\bot}}{(2\pi)^{d}}\frac{1}{{\bf p}_{\bot}^{2}+m^2}\exp(i{\bf p}_{\bot}.{\bf a})=\frac{1}{(2\pi)^{d/2}}m^{d-2}(ma)^{1-(d/2)}K_{(d/2)-1}(ma)\ ,
\end{equation}
and
\begin{equation}
\label{intdnot=2}
\int \frac{d^{d}{\bf p}_{\bot}}{(2\pi)^{d}}\frac{1}{{\bf p}_{\bot}^{2}}\exp(i{\bf p}_{\bot}.{\bf a})=\frac{1}{(2\pi)^{d/2}}2^{(d/2)-2}\Gamma\Biggl(\frac{d}{2}-1\Biggr)a^{2-d}\ d\not=2\ ,
\end{equation}
where $K$ stands for the $K$-Bessel function \cite{Arfken}, $\Gamma$ is the gamma function and $a=|{\bf a}|$.

When $d=2$ and $m=0$ we insert a regulator parameter, $\mu$, with mass dimension, as follows
\begin{eqnarray}
\label{intd=2}
\int \frac{d^{2}{\bf p}_{\bot}}{(2\pi)^{2}}\frac{1}{{\bf p}_{\bot}^{2}}\exp(i{\bf p}_{\bot}.{\bf a})&=&
\lim_{\mu\to0}\int \frac{d^{2}{\bf p}_{\bot}}{(2\pi)^{2}}\frac{1}{{\bf p}_{\bot}^{2}+\mu^2}\exp(i{\bf p}_{\bot}.{\bf a})=\frac{1}{(2\pi)}K_{0}(\mu a)\cr\cr
&=&-\frac{1}{(2\pi)}\Biggl[\ln\Biggl(\frac{ma}{2}\Biggr)+\gamma\Biggr]+\frac{1}{(2\pi)}\ln\Biggl(\frac{m}{\mu}\Biggr)\cr\cr
&\to& -\frac{1}{(2\pi)}\Biggl[\ln\Biggl(\frac{ma}{2}\Biggr)+\gamma\Biggr]
\end{eqnarray}
where in the second line we used the expansion $K_{0}(z)\stackrel{z\to 0}{\longrightarrow}-\ln(z/2)-\gamma$ ($\gamma$ stands for the Euler constant) and in the third line we discarded an $a$-independent term, which does not contribute to the interaction energy between the branes and, so, to the force between them. 

Now, using equations (\ref{intd=2}), (\ref{intdnot=2}) and (\ref{intm}) in (\ref{Eintermediario2}), so 
\begin{eqnarray}
\label{zxc3}
{\cal E}=\frac{1}{2}{\sum}_{k\neq s}\lambda_{k}\lambda_{s}\left\{
\begin{array}{ll}
\frac{1}{(2\pi)^{\frac{d}{2}}}
\biggl[2^{(\frac{d}{2}-2)}\Gamma(\frac{d}{2}-1)(a_{ks})^{2-d}-m^{d-2}(ma_{ks})^{1-(d/2)}K_{(d/2)-1}(ma_{ks})\biggr]\ d\neq2\cr
\frac{1}{(2\pi)}
\biggl[-\ln(\frac{ma_{ks}}{2})-\gamma-K_{0}(ma_{ks})\biggr]\ d=2
\end{array}
\right .
\end{eqnarray}

When we have two point-like branes in $3+1$ dimensions we must take $D=0$ and $d=3$ in (\ref{zxc3}). In this case we have the well known result obtained previously in the literature (see, for instance \cite{AcciolyPRD2004})
\begin{equation}
E(D=0,d=3)=\frac{\lambda_{1}\lambda_{2}}{4\pi}\frac{1-\exp(-ma)}{a}\ .
\end{equation}

The force density on the brane $a$ (force per unit of brane area) can be obtained by differentiation of (\ref{zxc3}), as follows
\begin{equation}
{{\cal F}_{a}}=-\sum_{b\not=a}\biggl(\frac{\partial\cal{E}}{\partial a_{ab}}\biggr)\frac{{\bf a}_{ab}}{a_{ab}}\ \ ,\ \ {\bf a}_{ab}={\bf a}_{a}-{\bf a}_{b},
\end{equation}
what leads to a single and general expression 
\begin{equation}
\label{zxc4}
{{\cal F}_{a}}=\sum_{b\not=a}\frac{{\lambda}_{a}{\lambda}_{b}}{(2\pi)^{\frac{d}{2}}(a_{ab})^{d-1}}
\biggl[2^{(\frac{d}{2}-1)}\Gamma\Biggl(\frac{d}{2}\Biggr)-(ma_{ab})^{\frac{d}{2}}K_{\frac{d}{2}}(ma_{ab})\biggr]\frac{{\bf a}_{ab}}{a_{ab}}\ \  d=1,2,3,...
\end{equation}

The results (\ref{zxc3}) and (\ref{zxc4}) show that the interaction between the sources are given by a Coulomb-like contribution with repulsive behavior (charges with the same signal repel each other) due to the massless modes and an Yukawa-like contribution with attractive behavior.

For completeness we point out that the interaction between dipole distributions along parallel $D$-dimensional branes can be obtained by using the same methods employed to compute (\ref{zxc3}). As discussed in reference \cite{BaroneHidalgo1}, the source which describes this kind of dipole distributions is given by
\begin{equation}
\label{Jdipolos}
J_{\mu}=\eta_{\mu0}\sum^{N}_{k=1}V^{\nu}_{(k)}\partial_{\nu}[\delta^{d}(\bf{x}_{\bot}-\bf{a}_{k})]
\end{equation}
where the the four vectors $V^{\nu}_{(k)}$ are taken to be constant and uniform in the reference frame we are performing the calculations.

Substituting (\ref{Jdipolos}) in (\ref{Egeral}) and performing the calculations analogously to the previous ones, one can show that
\begin{eqnarray}
{\cal E}=\frac{E}{L^{D}}&=&\sum^{N}_{k=1}{\sum^{N}_{s=1}}
\frac{2^{(\frac{d}{2}-2)}\Gamma(\frac{d}{2})}{(2\pi)^{\frac{d}{2}}{a_{ks}}^{d}}\biggr[d\biggr(\frac{{{\bf{V}}_{k}}_{\bot}.\bf{a}_{ks}}{a_{ks}}\biggl)\biggr(\frac{{{\bf{V}}_{s}}_{\bot}.\bf{a}_{ks}}{a_{ks}}\biggl)-{{\bf{V}}_{k}}_{\bot}.{{\bf{V}}_{s}}_{\bot}\biggl]+\cr\cr
&-&\frac{m^{d}}{(2\pi)^{\frac{d}{2}}}\biggr[(ma_{ks})^{-\frac{d}{2}}K_{\frac{d}{2}}(ma_{ks})\biggl({{\bf{V}}_{k}}_{\bot}.{{\bf{V}}_{s}}_{\bot}\biggr)\cr\cr
&-&(ma_{ks})^{-1-\frac{d}{2}}K_{1+\frac{d}{2}}(ma_{ks})\biggl({{\bf{V}}_{k}}_{\bot}.(m{\bf{a}}_{ks})\biggr)\biggl({{\bf{V}}_{s}}_{\bot}.(m{\bf{a}}_{ks})\biggr)\biggl]
\end{eqnarray}

To conclude this section we study the interaction energy between two point-like branes (what corresponds to $D=0$) at small distances and for higher dimensions. 

Taking the space-time dimensions as $1+1$, $2+1$, $3+1$, $4+1$ or $5+1$ we have respectively,
\begin{eqnarray}
\label{zxc5}
E(d=1)&=&-\lambda_{1}\lambda_{2}\Biggl(\frac{1}{2m}+\frac{m}{4}a^2+{\cal O}(a^{3})\Biggr)\cr\cr
E(d=2)&=&\frac{\lambda_{1}\lambda_{2}}{16\pi}\Biggl[\ln\Biggl(\frac{ma}{2}\Biggr)+(\gamma-1)\Biggr](ma)^2+{\cal O}[a^{4}\ln(ma)]\cr\cr
E(d=3)&=&\lambda_{1}\lambda_{2}\frac{m}{4\pi}\Biggl(1-\frac{1}{2}ma+O(a^2)\Biggr)\cr\cr
E(d=4)&=&-\lambda_{1}\lambda_{2}\frac{m^2}{16\pi^2}\Biggl[2\ln\Biggl(\frac{ma}{2}\Biggr)+(2\gamma-1)\Biggr]+{\cal O}[a^{2}\ln(ma)]\cr\cr
E(d=5)&=&\lambda_{1}\lambda_{2}\Biggl(\frac{m^2}{16\pi^2}\frac{1}{a}-\frac{m^3}{24\pi^{2}}+\frac{m^4}{64\pi^{2}}a+O(a^2)\Biggr)\ .
\end{eqnarray}

Notice that for $d=1,2,3$, the energies (\ref{zxc5}) are finite for $a=0$. For $d=4,5$ we have a divergent behavior for $a=0$, which is a general feature for $d\geq 4$. It is worth mentioning that the limit $a=0$ of the interaction energy between two point charges can be related to the self energy of a given point-like electric charge. This subject is not trivial and is under investigation \cite{Barone}. Some preliminary results indicate that, for higher dimensions, the self energy of a point charge may be rendered finite, via dimensional regularization, only for an odd number of spatial dimensions.

\section{Dirac string in the Lee-Wick electrodynamics}
\label{secaoDirac}

In this section we investigate the field strength produced by a Dirac string in the Lee-Wick electrodynamics. For this task we start by considering the field configuration of a Dirac string, lying on the $z$ axis, with magnetic flux $\Phi$ (positive along $\hat z$ direction), in the Maxwell electrodynamics
\begin{eqnarray}
\label{defAABM}
A^{\mu}_{Dirac(M)}(x)=
\frac{\Phi}{2\pi(x^{2}+y^{2})}(0,-y,x,0)
\end{eqnarray}
whose Fourier transform is \cite{Fernanda}
\begin{equation}
\label{Fo}
{\tilde{A}}^{\mu}_{Dirac(M)}(p)=(2\pi)^{2}\delta(p^{0})\delta(p^{3})\frac{i\Phi}{{\bf{p}}_{\bot}^{2}}(0,p_{y},-p_{x},0)\ ,
\end{equation}
where we defined the spatial perpendicular momentum to the string, ${\bf p}_{\bot}=(p_{x},p_{y},0)$ and the sub-index $M$ stands for quantities related to the Maxwell theory.

One can show that the external source of a Dirac string $J^{\mu}_{Dirac}(x)$, which produces the field (\ref{Fo}), has the Fourier transform \cite{Fernanda,Anderson}
\begin{equation}
\label{Je}
{\tilde{J}}^{\nu}_{Dirac}(p)=-p^{2}{\tilde{A}}^{\nu}_{Dirac,M}(p).
\end{equation}

The vector field produced by the Dirac string source (\ref{Je}) in the Lee-Wick electrodynamics is given by
\begin{eqnarray}
\label{Ah}
{A}^{\mu}_{Dirac}(x)&=&\int d^{4}y D^{\mu}_{\ \nu}(x,y)J_{Dirac}^{\nu}(y)\cr\cr
&=&\int \frac{d^{4}p}{(2\pi)^{4}}{\tilde{D}}^{\mu}_{\ \nu}(p){\tilde{J}}_{Dirac}^{\nu}(p)e^{-ip x}.
\end{eqnarray}
where the Lee-Wick propagator $D^{\mu\nu}(x,y)$ as well as its Fourier counterpart ${\tilde{D}}^{\mu\nu}(p)$ can be obtained from (\ref{propagador}). Taking the gauge parameter $\xi=1$ we have
\begin{equation}
\label{asd2}
{\tilde{D}}^{\mu\nu}(p)=\Biggl(\frac{1}{p^{2}-m^{2}}-\frac{1}{p^{2}}\Biggr)\left\{{\eta}_{\mu\nu}-\frac{p_{\mu}p_{\nu}}{m^{2}}\right\}\ .
\end{equation}

Using (\ref{asd2}), (\ref{Je}) and (\ref{Fo}) one can show that Eq. (\ref{Ah}) leads to
\begin{equation}
\label{asd5}
{A}^{\mu}_{Dirac}({\bf x}_{\bot})=i\Phi\int\frac{d^{2}\bf{p}_{\bot}}{(2\pi)^{2}}\Biggl(\frac{1}{{\bf{p}_{\bot}}^{2}}-\frac{1}{{\bf{p}_{\bot}}^{2}+m^{2}}\Biggr)(0,p_{y},-p_{x},0)\exp(i\bf{p}_{\bot}.\bf{x}_{\bot}).
\end{equation}
The first integral in (\ref{asd5}) is the vector potential produced by a Dirac string in Maxwell Electrodynamics (\ref{defAABM}), what can be verified with (\ref{Fo}). So, taking into account that the vector potential (\ref{asd5}) has only spatial components, we can write 
\begin{equation}
{\bf A}_{Dirac}({\bf x}_{\bot})={\bf A}_{Dirac(M)}({\bf x}_{\bot})+\Phi({\hat z}\times{\vec\nabla})\int\frac{d^{2}\bf{p}_{\bot}}{(2\pi)^{2}}\frac{1}{{\bf{p}_{\bot}}^{2}+m^{2}}\exp(i\bf{p}_{\bot}.\bf{x}_{\bot}),\ .
\end{equation}
From (\ref{intm}) one can get the result of the above integral and write
\begin{eqnarray}
\label{Adirac}
{\bf A}_{Dirac}({\bf x}_{\bot})&=&{\bf A}_{Dirac(M)}({\bf x}_{\bot})+\frac{\Phi}{2\pi}({\hat z}\times{\vec \nabla})K_{0}(m\sqrt{x^{2}+y^{2}})\cr\cr
&=&{\bf A}_{Dirac(M)}({\bf x}_{\bot})\Bigl[1-m\sqrt{x^{2}+y^{2}}K_{1}\Bigl(m\sqrt{x^{2}+y^{2}}\Bigr)\Bigr]\ ,
\end{eqnarray}
where, in the second line, we performed some simple manipulations.

Taking the rotational operator of (\ref{Adirac}) we have the magnetic field produced by a Dirac string in the Lee-Wick electrodynamics \footnote{We are neglecting the contribution for the magnetic field which come from ${\bf A}_{Dirac(M)}({\bf x}_{\bot})$. This contribution is divergent on the $z$ axis and vanishes in any other point of space.} 
\begin{equation}
\label{BLW}
{\bf B}_{Dirac}({\bf x}_{\bot})=\frac{\Phi}{2\pi}mK_{0}(m\sqrt{x^{2}+y^{2}})\hat{z}.
\end{equation}

It is interesting to notice that magnetic field (\ref{BLW}) points in the same direction of the magnetic flux inside the Dirac string. It diverges on the Dirac string and falls down quickly as we move away from it as
\begin{equation}
\label{BLW,mgrande}
{\bf B}_{Dirac}({\bf x}_{\bot}) \cong \frac{\Phi\ m}{2(2\pi)^{1/2}}\frac{\exp(-m\sqrt{x^{2}+y^{2}})}{(m\sqrt{x^{2}+y^{2}})^{1/2}}\hat{z},\ (m\sqrt{x^{2}+y^{2}}<<1)\ ,
\end{equation}
where we used the fact that $K_{0}(x)\cong [(2\pi)^{1/2}/2]\exp(-x)/x^{1/2}$ for $x>>1$.

Once we have an exterior magnetic field to the string it is natural to search for effects with no counterpart in the Maxwell theory. Let us start by taking two distinct Dirac strings. To simplify, we restrict to the case where they are parallel each other. One of them is taken to be lying on the $z$ axis and the other one, is parallel to the $z$ axis a distance $a$ apart. This system is described by the sources
\begin{eqnarray}
\label{fontesdirac}
J^{\mu}_{Dirac(1)}({\bf x})&=&\int\frac{d^{4}p}{(2\pi)^{4}}{\tilde J}^{\mu}_{Dirac}(p)e^{-ipx}\cr\cr
J^{\mu}_{Dirac(2)}({\bf x})&=&\int\frac{d^{4}p}{(2\pi)^{4}}{\tilde J}^{\mu}_{Dirac}(p)e^{-ipx}e^{i{\bf p}_{\bot}\cdot{\bf a}}
\end{eqnarray}
where ${\bf a}=(a_{x},a_{y},0)$ is a space vector for the string  position and ${\tilde J}^{\mu}_{Dirac}(p)$ is defined in (\ref{Je}).

Using the same arguments exposed in section (\ref{secaocargas}), one can show that the interaction energy between the two sources (\ref{fontesdirac}) is given by 
\begin{equation}
\label{intsole2a}
E=\frac{1}{T}\int d^{4}xd^{4}y J^{\mu}_{Dirac(1)}({\bf x})D_{\mu\nu}(x,y)J^{\nu}_{Dirac(2)}({\bf y})
\end{equation}
where we discarded contributions due to self interactions of the branes and used the fact that $D_{\mu\nu}(x,y)=D_{\mu\nu}(y,x)$.

It is more convenient to work in the momenta space. For this task we substitute (\ref{fontesdirac}) and (\ref{propagador}), with the gauge $\xi=1$, in (\ref{intsole2a}) and use definitions (\ref{asd2}), (\ref{Je}) and (\ref{Fo}), what leads to
\begin{eqnarray}
\label{iop1}
E&=&L\Phi_{1}\Phi_{2}\int\frac{d^{2}{\bf p}}{(2\pi)^{2}}\Biggl(\frac{1}{{\bf p}_{\bot}^{2}+m^{2}}-\frac{1}{{\bf p}_{\bot}^{2}}\Biggr){\bf p}_{\bot}^{2}e^{i{\bf p}_{\bot}\cdot{\bf a}}\cr\cr
&=&-L\Phi_{1}\Phi_{2}{\vec\nabla}_{\bf a}^{2}\int\frac{d^{2}{\bf p}}{(2\pi)^{2}}\Biggl(\frac{1}{{\bf p}_{\bot}^{2}+m^{2}}-\frac{1}{{\bf p}_{\bot}^{2}}\Biggr)e^{i{\bf p}_{\bot}\cdot{\bf a}}
\end{eqnarray}
where $L=\int dx^{3}$ is the string length and we defined the differential operator ${\vec\nabla}_{\bf a}^{2}=\frac{\partial^{2}}{\partial a_{x}^{2}}+\frac{\partial^{2}}{\partial a_{y}^{2}}$.

The above integrals are performed in (\ref{intdnot=2}) and (\ref{intd=2}). So, the energy (\ref{iop1}), after some simple manipulations, reads
\begin{eqnarray}
\frac{E}{L}=-\frac{m^{2}\Phi_{1}\Phi_{2}}{(2\pi)^{2}}K_{0}(ma)\ .
\end{eqnarray}
The corresponding force between the two solenoids is given by
\begin{equation}
\label{forcadirac}
F=-\frac{dE}{da}=-L\frac{\Phi_{1}\Phi_{2}}{(2\pi)^{2}}m^3K_{1}(ma).
\end{equation}
The force above is attractive if the two magnetic fluxes flow in the same direction, and repulsive otherwise. It is important to point out that this force falls down as fast as the distance between the solenoids, $a$, increases. This fact can be verified if one takes into account that the Bessel function $K_{1}(x)$ behaves like $K_{1}(x)\cong\exp(-x)/x^{1/2}$ for large $x$.

\section{Lee-Wick-like scalar model}
\label{secaoescalar}

For completeness, in this section we investigate some classical properties of a Lee-Wick-like model for the scalar field. To simplify, we restrict to $3+1$ dimensions. The corresponding lagrangian density is
\begin{equation}
\label{modeloescalar}
{\cal L}=\frac{1}{2}\partial_{\mu}\phi\partial^{\mu}\phi+\frac{1}{2}\partial_{\mu}\phi\frac{\partial_{\gamma}\partial^{\gamma}}{m^{2}}\partial^{\mu}\phi-\frac{1}{2}M^{2}\phi^{2}+J\phi
\end{equation}
with the corresponding propagator
\begin{equation}
\label{propescalar}
D(x,y)=\int\frac{d^{4}p}{(2\pi)^{4}}\frac{m^{2}}{p^{4}-m^{2}p^{2}+M^{2}m^{2}}\exp[-ip(x-y)]\ ,
\end{equation}
and dynamical equation
\begin{equation}
\Biggl(1+\frac{\partial_{\gamma}\partial^{\gamma}}{m^{2}}\Biggr)\partial_{\mu}\partial^{\mu}\phi+M^{2}\phi=J\ .
\end{equation}
In (\ref{modeloescalar}), $\phi$ is the scalar field and $J$ is an external source.

From (\ref{propescalar}) one can show that this model exhibits two massive poles for momentum square, namely
\begin{equation}
\label{defm+-}
m_{\pm}^{2}=\frac{m^{2}}{2}\Biggl(1\pm\sqrt{1-\frac{4M^{2}}{m^{2}}}\Biggr).
\end{equation}
In order to avoid tachyonic modes, one must take the restriction
\begin{equation}
0\leq\frac{4M^2}{m^{2}}\leq1\ .
\end{equation}
If $M=0$, we have a theory similar to the one studied in the previous section, for the electromagnetic field, with one massive mode, with mass $m$, and a massless one. This case is very similar to the one studied previously and has no novel physical properties.

If $0<4M^{2}/m^{2}<1$ we have two field modes with different non vanishing masses, $m_{+}$ and $m_{-}$, both of them lower than $m$ and $M$. In this case the propagator can be rewritten in the form 
\begin{equation}
D(x,y)=\int\frac{d^{4}p}{(2\pi)^{4}}\Biggl(\frac{1}{p^{2}-m_{+}^{2}}-\frac{1}{p^{2}-m_{-}^{2}}\Biggr)\frac{1}{\sqrt{1-\frac{4M^{2}}{m^{2}}}}\exp[-ip(x-y)]\ .
\end{equation}

It is interesting to consider the interaction energy between two time independent Dirac delta-like sources for the field concentrated at distinct points of space. This set-up is described by taking $J(x)=\sigma_{1}\delta^{3}({\bf x}-{\bf a}_{1})+\sigma_{2}\delta^{3}({\bf x}-{\bf a}_{2})$, where $\sigma_{1}$ and $\sigma_{2}$ are constants which, in some sense, stand for some kind of point-like charge for the scalar field. If one proceeds as in the previous section, one can show that the interaction energy between the sources is given by an attractive Yukawa-like potential with mass $m_{+}$ plus a repulsive Yukawa-like potential with mass $m_{-}$.
\begin{equation}
\label{escalardif}
E=\frac{\sigma_{1}\sigma_{2}}{4\pi}\frac{1}{\sqrt{1-\frac{4M^{2}}{m^{2}}}}\Biggl(\frac{\exp{(-m_{+}a)}}{a}-\frac{\exp{(-m_{-}a)}}{a}\Biggr)
\end{equation}

Once $m_{-}<m_{+}$, the energy (\ref{escalardif}) is always negative for $a\not=0$ and the interaction between the sources exhibits an attractive behavior, similarly to the Klein-Gordon field \cite{BaroneHidalgo1}, for which charges with the same signal attract each other. Notice that
in the  limit $a\to0$ the energy (\ref{escalardif}) is finite.

When $4M^{2}=m^{2}$ we have two massive poles for the momentum square at $m^{2}/2$. The propagator, now, reads  
\begin{equation}
D(x,y)=\int\frac{d^{4}p}{(2\pi)^{4}}\frac{m^{2}}{[p^{2}-(m^{2}/2)]^{2}}\exp[-ip(x-y)]\ ,
\end{equation}
and the interaction energy between the point-like sources is given by
\begin{equation}
\label{escalarig}
E=-\frac{\sigma_{1}\sigma_{2}}{4\pi}\frac{m}{\sqrt{2}}\exp{(-ma/\sqrt{2})}\ .
\end{equation}

In what concerns the above result we notice two interesting points. The first one is the fact that the energy (\ref{escalarig}) dominates for large distances in comparison with the single Yukawa potential, which would be obtained if we had used only one mass parameter in the model (\ref{modeloescalar}), $m$ or $M$. The second point is the attractive behavior of the interaction energy (\ref{escalarig}), {\it i.e}, charges with the same signal, $\sigma_{1}$ and $\sigma_{2}$, attract each other.

\section{Conclusions and Final Remarks}
\label{secaoconc}

In this paper we investigated some aspects of the interactions between stationary sources for fields in Lee-Wick-like abelian models. We studied the interactions between sources distributed along parallel branes with arbitrary dimension in the Lee-Wick electrodynamics in a space-time with an arbitrary number of spatial dimensions. We argued that, for point-like branes, the energy between them is finite when they are arbitrarily close to each other just for 1, 2 and 3 space dimensions. The obtained results indicate that for higher dimensions this energy diverges for charges infinitely close to each other. This fact deserves more investigations and it is an indication that the self energy of a point-like charge is finite, in Lee-Wick Electrodynamics, only for dimensions up to $3+1$.

Also in the context of Lee-Wick electrodynamics, we showed that a Dirac string, in $3+1$ dimensions, produces a magnetic field throughout the space which points in the same direction its internal magnetic flux flows. As an immediate consequence, we showed that two parallel stationary Dirac string in the Lee-Wick electrodynamics attract each other if their magnetic fluxes are in the same direction, and repel otherwise. The existence of a magnetic field exterior to the solenoid, as far as the authors know, were not explored in the literature since now. This fact can lead to other physical phenomena, like corrections to the Aharonov-Bohm effect, for instance, and is a subject which might be explored in other contexts. 

We studied a Lee-Wick-like model for the scalar field with two independent mass parameters. For simplicity, we considered only a $3+1$ spacetime, but the results can be generalized to an arbitrary number of spatial dimensions. By choosing conveniently the mass parameters, the interactions between point-like sources for the field can exhibit two distinct behaviors: a (double) Yukawa behavior or a single attractive exponential behavior, which falls down slower with the distance in comparison with the Yukawa interaction. In both cases the attractive behavior of the interactions still remains (scalar charges with the same signal attract each other) and the energy is finite if the distance between the charges is taken to be zero. 
These results may be important mainly in the context of the Lee-Wick Standard Model \cite{Grinstein2008}, in what concerns the Higgs field \cite{CaronePRD2009,CaroneJHEP2009}, and in the use of scalar Lee-Wick field in cosmological models \cite{Cai2009,ChoEPJC2013,ChoJCAP2011,LeeIJMP2008}. 

\ 

\noindent
{\bf Acknowledgments}

\noindent
F.A Barone, A.A.Nogueira and G. Flores-Hidalgo are very grateful, respectively, to CNPq, CAPES and FAPEMIG (Brazilian agencies) for financial support. The authors would like to thank J.A. Helay\"el-Neto and B.M. Pimentel for reading the paper suggestions to improve it.




\begin{thebibliography}{99}

\bibitem{Podolsky42} Boris Podolsky, Phys. Rev. D {\bf 62}, 68 (1942). 

\bibitem{Podolsky44} Boris Podolsy and Chihiro Kikuchi, Phys. Rev. D {\bf 65}, 228 (1944).

\bibitem{Podolsky48} Boris Podolsky and Philip Schwed, Rev. Mod. Phys. {\bf 29}, 40 (1948).

\bibitem{LW69} T.D. Lee and G.C. Wick, Nucl. Phys. {\bf B9}, 209 (1969).

\bibitem{LW70} T.D. Lee and G.C. Wick, Phys. Rev D {\bf 2}, 103 (1970).

\bibitem{Grinstein2008} Benjam\'\i n Grinstein, Donal O'Connell and Mark B. Wise, Phys. Rev. D {\bf 77}, 025012 (2008). 

\bibitem{Espinosa2008} Jos\'e Ram\'on Espinosa, Benjam\'\i n Grinstein, Donal O'Connell and Mark B. Wise, Phys. Rev. D {\bf 77}, 085002 (2008). 

\bibitem{Underwood2009} Thomas E.J. Underwood and Roman Zwicky, Phys. Rev. D {\bf 79}, 035016 (2009)

\bibitem{CaronePRD2009} Christopher D. Carone and Reinard Primulando, Phys. Rev. D {\bf 80}, 055020 (2009).

\bibitem{RizzoJHEP2008} Thomas G. Rizzo, JHEP {\bf 01}, 042 (2008).

\bibitem{RizzoJHEP2007} Thomas G. Rizzo, JHEP {\bf 06}, 070 (2007).

\bibitem{Schat2008} Ezequiel \'Alvarez, Carlos Schat, Leandro Da Rold and Alejandro Szynkman, JHEP {\bf 04}, 026 (2008).

\bibitem{KraussPRD2008} F. Krauss, T.E.E. Underwood and Zwicky, Phys. Rev. D {\bf 77}, 015012 (2008).

\bibitem{CuzinattoIJMPA2011} R.R. Cuzinatto, C.A.M. de Melo, L.G. Medeiros and P.J. Pompeia, Int. Jour. of Mod. Phys. A {\bf 26}, 3641 (2011).

\bibitem{AcciolyMPLA2010} A. Accioly and E. Scatena, Mod. Phys. Lett. A {\bf 25}, 269 (2010).

\bibitem{AcciolyMPLA2011} A. Accioly, P. Gaete, J.A. Helay\"el-Neto, E. Scatena and R. Turcati, Mod. Phys. Lett. A {\bf 26}, 1985 (2011).

\bibitem{Accioly2010} A. Accioly, P. Gaete, J.A. Helay\"el-Neto, E. Scatena and R. Turcati, Exploring Lee-Wick finite electrodynamics, (2010) [arXiv:physics/1012.1045v2].




\bibitem{Shalaby2009} Abouzeid M. Shalaby, Phys. Rev. D {\bf 80}, 025006 (2009).

\bibitem{CaronePLB2008} Christopher D. Carone and Richard F. Lebed, Phys. Lett. B {\bf 668}, 221 (2008).

\bibitem{CaronePLB2009} Christopher D. Carone, Phys. Lett. B {\bf 677}, 306 (2009).

\bibitem{CaroneJHEP2009} Christopher D. Carone and Richard F. Lebed, JHEP {\bf 01}, 043 (2009).

\bibitem{GrinsteinPRD2008} Benjam�n Grinstein and Donal O'Connell, Phys. Rev. D {\bf 78}, 105005 (2008).

\bibitem{gc} R. Sekhar Chivukula, Arsham Farzinnia, Roshan Foadi and Elizabeth H. Simmons, Phys. Rev. D {\bf 82}, 035015 (2010).

\bibitem{FornalPLB2009} Bartosz Fornal, Benjam\'i n Gristein and Mark B. Wise, Phys. Lett. B {\bf 674}, 330 (2009).

\bibitem{BoninPRD2010} C.A. Bonin, R. Bufalo, B.M. Pimentel and G.E.R Zambrano Phys. Rev. D {\bf 81}, 025003 (2010).

\bibitem{BoninPRD2011} C.A. Bonin and B.M. Pimentel Phys. Rev. D {\bf 84}, 065023 (2011).






\bibitem{Alekseev} A. I. Alekseev and B. A. Arbuzov, Theor. Math. Phys. {\bf 59}, 372 (1984); A. I. Alekseev, B. A. Arbuzov, and V. A. Baikov, Theor. Math. Phys. {\bf 52}, 739 (1982). 

\bibitem{BallNPB83} M. Baker, L. Carson, J.S. Ball and F. Zachariasen, Nucl. Phys. {\bf B229}, 456 (1983).

\bibitem{Baskal93} S. Baskal and T. Dereli, J. Phys. G: Nucl. Part. Phys. {\bf 19}, 477 (1993).

\bibitem{GOWPRD2008} Benjam\'\i n Grinstein, Donal O'Connell and Mark B. Wise, Phys. Rev. D {\bf 77}, 065010 (2008).


\bibitem{StellePRD77} K.S. Stelle, Phys. Rev. D {\bf 16}, 953 (1977).

\bibitem{StelleGRG77} K.S. Stelle, Gen, Rel. and Grav. {\bf 9}, 353 (1978).

\bibitem{WuPLB2008} Feng Wu and Ming Zhong, Phys. Lett. B {\bf 659}, 694 (2008).

\bibitem{WuPRD2008} Feng Wu and Ming Zhong, Phys. Rev. D {\bf 78}, 085010 (2008).

\bibitem{Rodigast2009} Andreas Rodigast and Theodor Schuster, Phys. Rev. D {\bf 79}, 125017 (2009).

\bibitem{Accioly2003} Antonio Accioly, Phys. Rev. D {\bf 67}, 127502 (2003).




\bibitem{Cai2009} Yi-Fu Cai, Taotao Qiu, Robert Brandenberger and Xinmin Zhang, Phys. Rev. D {\bf 80}, 023511 (2009).

\bibitem{ChoEPJC2013} Inyong Cho and O-Kab Kwon, Eur. Phys. Jour. C {\bf 73}, 2341 (2013).

\bibitem{ChoJCAP2011} I. Cho and O-K. Kwon, JCAP {\bf 043}, 1111 (2011).

\bibitem{LeeIJMP2008} Seokcheon Lee, Int. Jour. Mod. Phys.: Conference Series {\bf 1}, 252 (2008); arXiv:0810.1145[astro-ph](2008).




\bibitem{Frankel} J. Frankel, Phys. Rev. E {\bf 54}, 5859 (1996).

\bibitem{Zayats} Alexei E. Zayats, [arXiv:1306.3966] (2013).





\bibitem{SantosMPLA2011} Roberto Baginski and B. Santos, Mod. Phys. Lett. A {\bf 26}, 1909 (2011).

\bibitem{GabrielliPRD2008} Emidio Gabrielli, Phys. Rev. D {\bf 77}, 055020 (2008).


\bibitem{GalvaoCJP} Carlos A.P. Galv\~ao and B.M. Pimentel, Can. J. Phys. {\bf 66}, 460 (1987).

\bibitem{Bufalocan} M.C. Bertin, B.M. Pimentel and G.E. Zambrano, J.M. Phys. {\bf 52}, 102902 (2011).

\bibitem{BufaloPRD83} R. Bufalo, B.M. Pimentel and G.E.R. Zambrano, Phys. Rev. D {\bf 83}, 045007 (2011).


\bibitem{Moyaedi} S.K. Moyaedi, M.R. Setare and B. Khosropour, Adv. in H. En. Phys. {\bf 2013}, Article ID 657870 (2013).

\bibitem{GaeteJPA2012} Patr\'\i cio Gaete, Jos\'e Abdalla Helay\"el-Neto and Euro Spallucci, J. Phys. A: Math. Theor. {\bf 45}, 215401 (2012). 

\bibitem{CuzinattoAP2007} R.R. Cuzinatto, C.A.M. Melo and P.J. Pompeia, Ann. Phys. {\bf 322}, 1211 (2007).

\bibitem{KruglovJPA2010} S.I. Kruglov, J. Phys. A: Math. Theor. {\bf 43}, 245403 (2010). 

\bibitem{ChoPRD2010} I. Cho and O. K. Kwon, Phys. Rev. D {\bf 82}, 025013 (2010).



\bibitem{AcciolyPRD2004} Antonio Accioly and Marcos Dias, Phys. Rev. D {\bf 70}, 107705 (2004).

\bibitem{Zee} A. Zee, {\it Quantum Field Theory in a Nutshell}, Princeton University Press, (2003).

\bibitem{BaroneHidalgo1} F.A. Barone and G. Flores-Hidalgo, Phys. Rev. D {\bf 78}, 125003 (2008).

\bibitem{BaroneHidalgo2} F.A. Barone and G. Flores-Hidalgo, Braz. Jour. Phys. {\bf 40}, 188 (2010).


\bibitem{Arfken} G.B. Arfken and H.J. Weber, {\it Mathematical Methods for Physicists}, Academic Press (1995).



\bibitem{Barone} F.A. Barone {\it et al}, {\it The self energy of a point-like charge in Lee-Wick Electrodynamics}, work in progress. 

\bibitem{Fernanda} M.F.X.P. Medeiros, {\it Efeitos da Polariza\c c\~ao do V\'acuo nas Imedia\c c\~oes de um Solen\'oide}, Master's Thesis, Federal University of Itajub\'a (2012).

\bibitem{Anderson} A.A Nogueira, {\it Aspectos do modelo de Lee-Wick abeliano}, Master's Thesis, Federal University of Itajub\'a (2012). 

\end{thebibliography}
\end{document}